\documentstyle[12pt]{article}

\newcommand{\be}{\begin{equation}}
\newcommand{\ym}{Yang--Mills}
\newcommand{\ee}{\end{equation}}
\newcommand{\fu}{F^{ij}}
\newcommand{\fd}{F_{ij}}
\newcommand{\di}{D_i}
\newcommand{\eps}{\epsilon_{ijk}}
\newcommand{\bes}{\begin{eqnarray}\displaystyle}
\newcommand{\ees}{\end{eqnarray}}
\newcommand{\p}{\partial}
\newcommand{\non}{\nonumber}

\newcommand{\half}{{1\over 2}}
\newcommand{\quater}{{1\over 4}}

\newcommand{\teu}{{\rm \bf R}^3}

\newcommand{\refb}[1]{(\ref{#1})}

\def\dash{D\!\!\!\! / \,}
\def\pash{\partial\!\!\! / \, }
\def\bash{B\!\!\!\! / \,}

\begin{document}

{}~ \hfill\vbox{\hbox{{\sf hep-th/9702116}}\hbox{DAMTP/97-9}
}\break

\vskip 3.5cm
\begin{center}

{\large\bf Non-confinement in Three Dimensional} \\
\vskip 0.2cm
{\large\bf  Supersymmetric Yang--Mills Theory}

\vskip 1.5cm

{Hwang-hyun Kwon}

\vskip 0.2cm
{\it Department of Applied Mathematics and Theoretical Physics \\
University of Cambridge \\ Cambridge CB3 9EW, UK  }

{\sf h.kwon@amtp.cam.ac.uk}

\vskip 2.0cm

\abstract{The role of instantons in three dimensional N=2
supersymmetric $SU(2)$ Yang--Mills theory is studied, especially in relation
to the issue of confinement. The instanton-induced low energy
effective action is derived
by extending the dilute gas approximation to the super-moduli space of
instantons. Following Polyakov's description of confinement in compact
$U(1)$ gauge theory, it  is argued that there
is no confinement in N=2 supersymmetric Yang--Mills theory.}

\end{center}

\vfill

\baselineskip=16pt

\section{Introduction}

Confinement in three dimensional compact $U(1)$ gauge theory was
demonstrated analytically by Polyakov \cite{poly1} in 1977. He showed
that the Wilson loop has an area dependence which arises from
instanton effects, where the instanton
is the 't Hooft--Polyakov monopole. The
photon becomes massive by instanton condensation so a mass gap is generated. 

The main purpose of this work is to extend the analysis to a 
supersymmetric $SU(2)$ Yang--Mills theory without matter multiplets in
which the instanton is a BPS monopole. Since monopole solutions
require Higgs fields, we need, at least, N=2 supersymmetry. BPS monopoles 
preserve half the supersymmetry and the broken supersymmetry generates
fermionic zero modes. This is also true when the monopoles are
interpreted as instantons in three dimensions.
These configurations also have very particular dynamics compared to generic
't Hooft--Polyakov monopoles \cite{manton}.   
In the BPS limit of the bosonic theory, this leads to a
somewhat singular limit of Polyakov's considerations. We will see that
a close relation emerges between the instanton-induced low energy
effective action and a complex Toda action \cite{evans} in three dimensions. 

In the N=2 supersymmetric theory, the low energy effective action can
be written in terms of a photon, a Higgs scalar and a complex fermion
field \cite{affleck}. This action was obtained in 
\cite{affleck} from symmetry arguments and calculating instanton
contributions to a fermion propagator. But 
it can be explicitly derived by extending   
the dilute gas approximation to
the superspace consisting of collective coordinates of
BPS monopoles. 
Such a  dilute gas approximation is more appropriate for
a monopole gas than the BPST instanton gas in four dimensions since
the size of monopoles is fixed by the parameters of the theory.    
It can be observed from the effective
action that there is no mass gap in the supersymmetric theory, which
could be a sign of non-confinement. Indeed, we will see by the Wilson
loop criterion
that there is no confinement in N=2 supersymmetric \ym\ theory. 

The low energy effective action of the bosonic theory in the BPS limit
is given in section 2 and the question of confinement is discussed. The
supersymmetric case is discussed in section 3. Some features of the
BPS monopole as an instanton solution of the three dimensional system
are reviewed in Appendix.

\section{The BPS limit of the bosonic theory} 

Before embarking on the supersymmetric theory, we will discuss the BPS
limit of Polyakov's arguments in \cite{poly1}, where he studied $SU(2)$ 
Yang--Mills--Higgs theory spontaneously broken to $U(1)$. 
The partition function of interest is given by, in $\teu$,
\be
{\cal Z} = \int {\cal D}A_i{\cal D}\Phi\; {\rm exp}\Big[-\int d^3 x\;
Tr\Big({1\over 4}\fd\fu + \half \di^2\Phi\Big)\Big]    
\ee
where $\Phi$ is the Higgs field in the adjoint representation of
$SU(2)$ and $i, j = 1,2,3$.

The action for a multi-instanton
configuration is the energy of the corresponding static
multi-BPS monopole configuration in four dimensions, which gives, in a
dilute gas approximation,
\be\label{zall}
{\cal Z} = \sum_{N,q(=\pm 1)}{\zeta^N\over N!}\int\prod_{i=1}^N d^3{\rm R}_i
{\rm exp}\Big(-{\pi\over 2e^2}\sum_{i\not= j}^N{q_iq_j-1\over {|{\rm
R}_i-{\rm R}_j}|} \Big) 
\ee
where $\zeta = c\: e^{-4\pi v/ e}$, $v$ is the vacuum expectation
value of $\Phi$ and $c$ is the one-loop
functional determinant in a single BPS monopole background. The arguments
in the exponent account for the fact that there is no force between
like charges and the attraction between opposite
charges is double the Coulomb force. These
multi-BPS monopole configurations satisfy the Bogomol'nyi equation only
when they consist of like charges. But the mixed combination of
BPS monopoles and anti-BPS monopoles are still good approximate
solutions of the equations of motion. 

The partition function can be written, using
Gaussian integral identities, as a functional integral over two real
scalar fields, representing the photon and the Higgs scalar,
\bes\label{eff1}
{\cal Z} &=& C\sum_{N,q(=\pm 1)}{\zeta^N\over N!}\int\prod_{i=1}^N
d^3{\rm R}_i\int {\cal D}\gamma{\cal D}\phi\;e^{-({e\over 2\pi})^2\int d^3x
\{\half(\nabla\gamma)^2+\half(\nabla\phi)^2\}+ i\sum_{i=1}^{N}q_i\gamma({\rm
R}_i)- \sum_{i=1}^{N} \phi({\rm R}_i)} \non \\
&=&C\int {\cal D}\gamma{\cal D}\phi\;e^{-({e\over 2\pi})^2\int d^3x
\{\half(\nabla\gamma)^2+\half(\nabla\phi)^2\}}\sum_N{\zeta^N\over
N!}\int\prod_{i=1}^N \Big(d^3{\rm R}_i\;2\cos \gamma ({\rm
R}_i)e^{-\phi ({\rm R}_i)}\Big) \non \\
&=& C\int {\cal D}\gamma{\cal D}\phi\;{\rm exp}\Big[-({e\over2\pi})^2
\int d^3x\Big\{\half(\nabla\gamma)^2+\half(\nabla\phi)^2-{m^2\over 2}
(e^{-\phi+i\gamma}+e^{-\phi-i\gamma})\Big\}\Big]   
\ees
where $C={\rm det}|-{e^2\over2\pi^2}\nabla^2|$ and $m^2=2\zeta ({2\pi\over e})^2 $. The instanton-induced effective
action in \refb{eff1} is a complex Toda-like action \cite
{evans} in three dimensions. It has a rather singular behaviour caused
by the $e^{-\phi}$ term which makes the system unstable. But
the effective action will be useful for comparison with that of the
supersymmetric theory to be considered in the next section.   
 
The Wilson loop can be calculated, following
 \cite{poly1}, and is given in
the semi-classical approximation as
\bes\label{swil}
W &\equiv& \Big<\:{\rm exp}\Big(i\oint_{{\mit\Gamma}} A^3_i d
x^i\Big)\:\Big> = \Big<\;{\rm exp}\Big(i\int d^3x\:\rho({\bf
x})\eta({\bf x})\Big)\;\Big> \non \\ 
&=&{\rm exp}\Big[-({e\over2\pi})^2\int d^3x \Big\{
\half(\nabla\gamma_c-\nabla\eta)^2+\half(\nabla\phi_c)^2  -m^2\cos \gamma_c
\: e^{-\phi_c}\Big\}\Big]
\ees
where the loop $\Gamma$ is located in the $x_1$-$x_2$ plane, $\rho$ is
the monopole charge density and $\eta$ is defined by 
\be\label{eta}
\eta ({\bf x}) = \half \int_S d {\bf S}_y^i {({\bf y}-{\bf x})^i\over {|{\bf
y}-{\bf x}|^3}} \quad .
\ee
The fields $\gamma_c, \phi_c$ satisfy
\bes\label{deq}  
\nabla^2\gamma_c &=& \nabla^2\eta +m^2\sin\gamma_c\: e^{-\phi_c} \non \\
&=& -2\pi\delta '(x_3)\theta_S(x_1,x_2)
+m^2\sin\gamma_c\: e^{-\phi_c} \non \\
\nabla^2\phi_c &=& m^2 \cos\gamma_c\: e^{-\phi_c}
\ees
where $S$ is the minimal surface whose boundary is the loop $\Gamma$ and
\bes\label{thetas}
\theta_S(x_1,x_2) &=& 1 \qquad{\rm if}\quad (x_1,x_2) \in S\, , \non \\
&=& 0 \qquad {\rm otherwise}\, . 
\ees
For a large loop with order of $R^2$, the system is essentially
one-dimensional for $x_1^2+x_2^2 \ll R^2$ and \refb{deq} can be 
reduced to
\be\label{toda} 
{d^2\gamma_c\over dx_3^2} = m^2 \sin \gamma_c\: e^{-\phi_c}, \qquad 
{d^2\phi_c\over dx_3^2} = m^2 \cos \gamma_c\: e^{-\phi_c}\, .
\ee
Eq.\refb{toda} should be solved to check confinement but
 an analytic solution has not been found. For our purposes, this is
only part of the complete supersymmetric theory to which we now turn.

\section{ N=2 Supersymmetric \ym\ theory}

Inclusion of dynamical fermions can cause problems for the the Wilson
loop criterion of confinement \cite{bander}. But the dynamical
fermions in supersymmetric \ym\
theories are in the adjoint representation of the gauge group, whereas
the test charges are in
the fundamental representation.     

The partition function of N=2 supersymmetric Yang--Mills
theory in Euclidean space is  
\be\label{part}
{\cal Z} =\int{\cal D}A_i{\cal D}\Phi{\cal D}\Psi^*{\cal D}
\Psi {\rm exp}\Big[-\int d^3 x\; Tr\Big(\:\quater\fd\fu +\half \di^2\Phi 
+\Psi^{\dagger}(i\dash +e[\Phi,\, ])\, \Psi\:\Big)\Big].
\ee
where $\Psi, \Psi^*$ must be treated as independent Dirac spinors.
The supersymmetry transformations are
\bes\label{susytr}
\delta A_i &=& i\Psi^{\dagger}\sigma_i\alpha
-i\alpha^{\dagger}\sigma_i\Psi \non \\ 
\delta\Phi &=& i\Psi^{\dagger}\alpha -i\alpha^{\dagger}\Psi \non \\
\delta\Psi &=& i(\bash-\dash\Phi)\alpha, \qquad     
\delta\Psi^* =  -i(\bash +\dash\Phi)\alpha^*\,\, .
\ees

In addition to the translational zero modes, instantons now have fermionic
zero modes  generated by the
supersymmetry transformations \refb{susytr}. These fermionic zero modes
need to be included in the
effective action calculation. The superfield formalism turns out to be
efficient for this purpose \cite{shiff}. 
Collective coordinates
of anti-instantons and instantons will be denoted as $({\rm
x}_i,\alpha_i)$ and $({\rm y}_i,\alpha^*_i)$, respectively. The
instanton superfield $V_S$ with a collective
coordinate $({\rm y},\alpha^*)$ can be written as
\be
V_S({\rm y},\alpha^*) = e^{-i{\rm y}\cdot P+i\alpha^*Q^*}V_B({\rm y}=0)
\ee
where $V_B$ represents the bosonic instanton
configuration. Under the complex supersymmetry transformation with an
algebra $\{Q,Q^*\}= 4\,{\bf\sigma}\!\cdot\! P$, 
\bes\label{sfil}
V_S &\rightarrow & e^{i\epsilon Q+i\epsilon^*Q^*}e^{-i{\rm y}\cdot
P+i\alpha^*Q^*}V_B  \non \\
&=& {\rm exp}\{-i\,({\rm y}-2i\epsilon{\bf \sigma}\alpha^*)\!\cdot\!
P+i\,(\alpha^*+\epsilon^*)Q^*\}V_B
\ees
where ${\bf \sigma}=(\sigma_1,\sigma_2,\sigma_3)$.
Hence, supersymmetry transformations induce translations of instanton
collective coordinates as 
\be\label{collec1}
\delta {\rm y}_j^a = -2i\epsilon\sigma^a\alpha_j^* ,\qquad
\delta\alpha^*_j = \epsilon^* \,\, .\qquad (a=1,2,3)
\ee
Those for anti-instantons can be obtained in the same way, which are
\be\label{collec2}
\delta {\rm x}_i^a = 2i\alpha_i\sigma^a\epsilon^* , \qquad
\delta\alpha_i = \epsilon \,\, .
\ee
 
Now the instanton contribution can be represented by the partition
function of the dilute instanton gas in the superspace consisting of
their collective coordinates \cite{yung}. The remaining task is to
supersymmetrize the Coulomb potential of the bosonic theory.
The supersymmetry-invariant distance between $({\rm
x}_i,\alpha_i)$ and $({\rm y}_i,\alpha^*_i)$ is 
\be
|{\rm x}_i-{\rm y}_j-2i\alpha_i{\bf \sigma}\alpha_j^*| 
\ee
which is easy to check using \refb{collec1} and \refb{collec2}.        
The partition function \refb{part} can, then, be written as
\bes\label{scs}
{\cal Z} &=& \sum_{M,N}{\zeta^{M+N}\over M!N!}\int\prod_{i=1}^M d^3{\rm x}_i
d^2\alpha_i \prod_{j=1}^N d^3{\rm y}_j d^2\alpha^*_j 
{\rm exp}\Big({\pi\over e^2}\sum_{i,j}^{M,N}{1\over {|{\rm x}_i-{\rm
y}_j-2i\alpha_i{\bf\sigma}\alpha_j^*}|} \Big) \non \\ 
&=& \sum_{M,N}{\zeta^{M+N}\over M!N!}\int\prod_{i=1}^M d^3{\rm x}_i
d^2\alpha_i \prod_{j=1}^N d^3{\rm y}_j d^2\alpha^*_j 
{\rm exp}\Big({\pi\over e^2}\sum_{i,j}^{M,N}{1\over {|{\rm x}_i-{\rm
y}_j|}}\Big( 1 +2i\,\Big|{{\alpha_i{\bf\sigma}\alpha_j^*}\over {{\rm x}_i-{\rm
y}_j}}\Big|\, \Big) \non \\ 
&=& \sum_{M,N}{\zeta^{M+N}\over M!N!}\int\prod_{i=1}^M d^3{\rm x}_i
d^2\alpha_i \prod_{j=1}^N d^3{\rm y}_j d^2\alpha^*_j 
{\rm exp}\Big({\pi\over e^2}\sum_{i,j}^{M,N}e^{-2i\alpha_i
{\bf \sigma}\alpha^*_j{\p\over \p {\rm x}_i}} {1\over {|{\rm x}_i-{\rm y}_j}|} \Big) 
\ees
where the potential energy comes only from the oppositely
charged BPS monopoles.

As in the bosonic theory, the exponent can be
expressed as a functional integral, this time, over a N=2 scalar superfield,
\bes\label{scs1}
{\cal Z} &=& \sum_{M,N}{\zeta^{M+N}\over M!N!}\int\prod_{i=1}^M d^3{\rm x}_i
d^2\alpha_i \prod_{j=1}^N d^3{\rm y}_j d^2\alpha^*_j \int{\cal
D}{\mit \Phi}^* {\cal D}{\mit \Phi} \;{\rm exp}\Big[\int d^3 x
d^2\theta d^2\theta^*\, {\mit \Phi}^*{\mit \Phi} \non \\
&&-\sum_{i=1}^{M} {2\pi\over e}\; {\mit \Phi}({\rm x}_i,\alpha_i) 
-\sum_{j=1}^{N} {2\pi\over e}\; {\mit \Phi}^*({\rm y}_j,\alpha^*_j)\;\Big]    \non \\
&=& \!\!\int{\cal D}{\mit \Phi}^* {\cal D}{\mit \Phi}\;{\rm
exp}\Big[\int d^3 x \Big\{ \int d^2\theta d^2\theta^*\, {\mit
\Phi}^*{\mit \Phi}+\int d^2\theta \: {\cal W}({\mit \Phi}) +\int
d^2\theta^*\: {\cal W}({\mit \Phi}^*)\Big\}\Big]  
\ees
where the non-perturbative superpotential 
\be\label{spot}
{\cal W}({\mit \Phi})= \zeta \;{\rm exp}\:\Big(-{2\pi\over e}\:{\mit \Phi}\Big) .
\ee
As can be expected from the close relation between the effective
action of the bosonic theory and the complex Toda action, the action
in \refb{scs1} is that of the N=2
supersymmetric $A_1$ Toda theory in three dimensions \cite{sevans}.

Defining ${\mit \Phi} \equiv Z +
\sqrt{2}\theta\psi +\theta\theta F$ and integrating out the
auxiliary field $F$, the instanton-induced effective lagrangian can be
recast in terms of component fields as
\be\label{seff2} 
{\cal L} = \p_iZ^*\p^iZ +
i\psi^{\dagger}\pash\psi +\half\zeta \Big({2\pi\over e}\Big)^2\Big(\psi\psi
e^{-{2\pi\over e}Z}+ \psi^*\psi^* e^{-{2\pi\over e}Z^*}\Big)
+\Big({2\pi\over e}\zeta\Big)^2 e^{-{2\pi\over e}(Z +Z^*)}  \non \\
\ee 
where ${\cal L}$ is defined by
\be
{\cal Z}= \int {\cal D}^2Z{\cal D}\psi^*{\cal D}\psi{\cal D}^2F\; {\rm
exp}\Big(-\int d^3 x \;{\cal L}\Big)\,\, .
\ee
Thus, the dilute gas approximation in superspace has enabled us to derive
the bosonic potential term in \refb{seff2}  which was motivated 
in \cite{affleck} by requiring the effective
action to be supersymmetric. The potential terms in \refb{eff1} are now
coupled with the fermionic terms because of the fermionic zero-modes
of BPS monopoles.

Following \refb{swil}, the Wilson loop is  
\bes\label{suwil}
W &=& \Big< e^{i\int d^3 x\:\rho(x)\eta(x)}\Big > \non \\
&=& \sum_{M,N}{\zeta^{M+N}\over M!N!}\int\prod_{i=1}^M d^3{\rm x}_i
d^2\alpha_i \prod_{j=1}^N d^3{\rm y}_j d^2\alpha^*_j \int{\cal
D}{\mit\Phi}^* {\cal D}{\mit\Phi} \;{\rm exp}\Big[\int d^3 x
d^2\theta d^2\theta^*\, {\mit\Phi}^*{\mit\Phi} \non \\
&&- \sum_{i=1}^{M} \Big\{ {2\pi\over e} {\mit\Phi}({\rm
x}_i,\alpha_i)+i\eta({\rm x}_i)\Big\}-\sum_{j=1}^{N} \Big\{ {2\pi\over e} {\mit
\Phi}^*({\rm y}_j,\alpha^*_j)-i\eta({\rm y}_j)\Big\}\;\Big] \non \\    
&=& \int {\cal D}^2 {\mit\Phi}\;{\rm
exp}\Big[\int d^3 x \Big\{ \int d^2\theta d^2\theta^*\, {\mit\Phi}^* {\mit\Phi}+\int d^2\theta \: e^{-{2\pi\over e}{\mit\Phi}-i\eta} +\int
d^2\theta^*\: e^{-{2\pi\over e}{\mit\Phi}^*+i\eta}\Big\}\Big]
\ees
In terms of component fields, \refb{suwil} can be expressed as
\bes\label{fwil}
W &=& \int{\cal D}^2 Z{\cal
D}\psi^*{\cal D} \psi\; {\rm exp}\Big[-\int d^3
x\;\Big\{\:\Big( \nabla Z -i{e\over 2\pi} \nabla\eta \, \Big)^2  \non \\
&&+\half\zeta \Big({2\pi\over e}\Big)^2\Big(\psi\psi
e^{-{2\pi\over e}Z}+ \psi^*\psi^* e^{-{2\pi\over e}Z^*}\Big)
+\Big({2\pi\over e}\zeta\Big)^2 e^{-{2\pi\over e}(Z +Z^*)}  \Big\}\Big]\,\, .  
\ees

Integrating out $\psi,\psi^*$ gives an effective potential
$f\Big(e^{-{2\pi\over e}(Z +Z^*)}\Big)$ whose explicit form will be
unimportant for the discussion. It is important that
$f$ depends only on the real part of the complex scalar field $Z$.
Therefore, we can write \refb{fwil} as
\be
W = \int{\cal D}\gamma{\cal D}\phi\; {\rm exp}\Big[-\int d^3
x\;\Big\{\:\half \Big(\nabla\gamma-{e\over \sqrt{2}\pi}\nabla\eta
\Big)^2++\half(\nabla \phi)^2 +U(\phi)\Big\}\Big]
\ee
where $\phi$ and $\gamma$ are defined by
$Z={1\over\sqrt{2}}(\phi+i\gamma)$ and $U(\phi)= \Big({2\pi\over e}\zeta\Big)^2 e^{-{2\sqrt{2}\pi\over e}\phi}+f(\phi)$.  
In the semi-classical approximation, \refb{fwil} is given by 
\be\label{fwil1}
W = {\rm exp}\Big[-\int d^3x \Big\{
\half \Big(\nabla\gamma_c-{e\over \sqrt{2}\pi}\nabla\eta \Big)^2+\half(\nabla\phi_c)^2+ U(\phi_c)\Big\}\Big]
\ee
where $\gamma_c$ and $\phi_c$ satisfy
\bes\label{sdeq}  
\nabla^2\gamma_c &=& {e\over \sqrt{2}\pi}\nabla^2\eta \non \\
\nabla^2\phi_c &=& {dU\over d\phi_c} \, .
\ees  
 Now it can be seen from
\refb{sdeq} that $\phi_c$ is completely
decoupled from the source term. It will only give some overall
numerical factor. The classical field $\gamma_c$ is analogous to the electric
potential due to a dipole layer in electrodynamics \cite{wadia}.
By partial integration, \refb{fwil1} can be written as
\be
W = N\, {\rm exp}\Big[\half\int d^3x 
\Big(\gamma_c-{e\over \sqrt{2}\pi}\eta\Big)\Big(\nabla^2\gamma_c-{e\over \sqrt{2}\pi}\nabla^2\eta\Big)\Big]
\ee
where $N$ represents various numerical factors. 
From \refb{sdeq}, the argument in the exponent vanishes and there is no
area law behaviour. The fluctuation around the classical configuration
was argued in \cite{wadia} to give the Wilson loop a perimeter
dependence. Hence, there is no confinement in 
 N=2 supersymmetric Yang--Mills theory. This could also be
anticipated from the effective action \refb{seff2}, where instanton
effects make the fermions massive but the photon field $\gamma$
remains massless.

\vskip 0.3cm

In summary,  we have derived the instanton-induced effective action of the
N=2 supersymmetric Yang--Mills theory in a manner that parallels the
discussion of the non-supersymmetric theory in \cite{poly1}. In the
presence of the extended supersymmetry, the instanton gas does not
lead to confinement.
The essential feature is that the non-perturbative
potential which makes a photon massive in the bosonic theory does not occur
in the supersymmetric case. Instanton effects generate mass
terms only for fermions. In \cite{sw3}, the N=4
supersymmetric Yang--Mills theory without matter
multiplets was also argued to have no confinement.

\vskip 0.8cm

{\bf Acknowledgements}: I would like to thank Michael Green for
suggesting the problem, insightful discussions and encouragements. I
also appreciate discussions with Jonathan Evans, Michael Gutperle and
Nick Manton on some issues in the paper. This work was supported by
COT, ORS Award and Dong-yung Scholarship.
  
\vskip 3.5cm

\noindent{\large {\bf Appendix :$\;\;$ BPS monopoles as instantons in
three dimensions}}

\vskip 0.3cm

't Hooft--Polyakov monopoles arise as instanton solutions of the three 
dimensional  $SU(2)$ Yang--Mills--Higgs theory spontaneously broken to
$U(1)$ whose action in $\teu$ is
\be\label{ggaction} 
{\cal S} = \int d^3x\; Tr \Big(\;{1\over 4} \fd \fu + {1\over 2} 
D^2_i \Phi +{\lambda\over 4} (\Phi^2 - a^2)^2\;\Big)\,\, .
\ee
It can be easily seen that, in the BPS limit $(\lambda=0)$, the action is minimised
by the field configuration satisfying
\be \label{bogo}
B_i = \pm \di \Phi\, ,    
\ee
where we defined $B_i={1\over 2} \eps F^{jk}$, the dual of
$\fd$. Eq.\refb{bogo} is known as the Bogomol'nyi equation whose
solution is the BPS monopole. 

Although these instantons are simply BPS monopole configurations in four
dimensions,
they have certain features
particular to three dimensions. Since there is no
extra time coordinate on which the fields may depend, there is no
analogue of the dyon solutions of four dimensions which 
depend on time. In other words, there is no $A_0$ field that
characterises the dyon solutions \cite{julia}. We can ask, however,
whether they have an electric
charge in Euclidean sense, an unbroken $U(1)$ charge. This $U(1)$ gauge transformation is the rotation around $\Phi /
|\Phi|$ \cite{witten}. Under the infinitesimal gauge transformation, the
fields transform as
\be
\delta \Phi = 0, \qquad \delta A_i = - {1\over e v } \di \Phi
\ee
where $v$ is the vacuum expectation value of $\Phi$. 
In Euclidean space, the Noether current is 
\be
J^i = {1\over ev} Tr \Big(\:\eps B^j D^k \Phi \:\Big)
\ee  
hence, by the Bogomol'nyi equation \refb{bogo}, the $U(1)$ charge vanishes
for BPS monopoles. But non-BPS monopoles can have non-zero $U(1)$
charge. There is no contribution from a $\theta$-term to the $U(1)$
current since the $\theta$-term of the theory is
\be\label{theta}
{\cal L}_\theta = i{e\theta\over 4\pi v} Tr\Big(\: B^iD_i\Phi \:\Big)
\ee
and its contribution to the current $J^i$ is
\bes
J^i_\theta &=& i{\theta\over 4\pi v^2}\epsilon^{ijk}\;Tr\Big(\: D_j\Phi D_k\Phi\:\Big)
\non \\ &=& 0 \,\, .
\ees

\end{document}